\documentclass[]{aa}
\usepackage{times}
\usepackage{natbib}
\usepackage[dvips]{graphicx}
\usepackage{epsfig,longtable,lscape}
\usepackage{epsfig}
\usepackage{rotating}
\bibpunct[]{(}{)}{;}{a}{}{,}

\def\XMM{{\em XMM--Newton}}

\def\Chandra{{\em Chandra}}
\def\pn{{\em pn}}
\def\BDone{BD\,+37$^\circ$\,442}   
\def\BDtwo{BD\,+37$^\circ$\,1977}   
\def\BDthree{{BD+28$^\circ$ 4211}}

\def\HD{HD\,49798}
\def\approxgt{\mathrel{\hbox{\rlap{\lower.55ex \hbox {$\sim$}}
        \kern-.3em \raise.4ex \hbox{$>$}}}}
\def\approxlt{\mathrel{\hbox{\rlap{\lower.55ex \hbox {$\sim$}}
        \kern-.3em \raise.4ex \hbox{$<$}}}}

\def\ltsima{$\; \buildrel < \over \sim \;$}
\def\lsim{\lower.5ex\hbox{\ltsima}}
\def\gtsima{$\; \buildrel > \over \sim \;$}
\def\gsim{\lower.5ex\hbox{\gtsima}}

\begin{document}

\title{Follow-up observations of X-ray emitting hot subdwarf star: the He-rich sdO \BDtwo}

\author{N. La Palombara\inst{1}, P. Esposito\inst{1,2}, S. Mereghetti\inst{1}, G. Novara\inst{1,3}, A. Tiengo\inst{1,3,4}}

\institute{INAF, Istituto di Astrofisica Spaziale e Fisica Cosmica - Milano, via Bassini 15, I-20133, Milano, Italy
\and Harvard-Smithsonian Center for Astrophysics, 60 Garden Street, Cambridge, MA, 02138, USA
\and IUSS-Istituto Universitario di Studi Superiori, piazza della Vittoria 15, I-27100 Pavia, Italy
\and Istituto Nazionale di Fisica Nucleare, Sezione di Pavia, via A. Bassi 6, I-27100 Pavia, Italy}

\titlerunning{\XMM\ observation of \BDtwo}

\authorrunning{La Palombara et al.}

\abstract{
We report on the results of the first \XMM\ satellite observation of the luminous and helium--rich O--type subdwarf \BDtwo\ carried out in April 2014. X--ray emission is detected with a flux of about 4$\times10^{-14}$ erg cm$^{-2}$ s$^{-1}$ (0.2--1.5 keV), corresponding to a $f_{\rm X}/f_{\rm bol}$ ratio $\sim 10^{-7}$; the source spectrum is very soft, and is well fit by the sum of two plasma components at different temperatures. Both characteristics are in agreement with what is observed in the main-sequence early-type stars, where the observed X-ray emission is due to turbulence and shocks in the stellar wind. A smaller but still significant stellar wind has been observed also in \BDtwo; therefore, we suggest that also in this case the detected X-ray flux has the same origin.
}

\maketitle

\section{Introduction}\label{introduction}

\begin{table*}[t]
\caption{Main parameters of the sdO stars \BDtwo, \BDone, and \HD.}\label{parameters}
\begin{center}
\begin{tabular}{lc|cc|cc|cc} \hline \hline
Parameter				& Symbol	& \multicolumn{2}{c}{\BDtwo}	& \multicolumn{2}{c}{\BDone}		& \multicolumn{2}{c}{\HD}		\\
					&		& Value		& Reference	& Value			& Reference	& Value			& Reference	\\ \hline
Surface gravity				& log $g$	& $\simeq$ 4.0	& 1		& 4.00 $\pm$ 0.25	& 4		& 4.35			& 6		\\
Luminosity ($L_{\odot}$)		& $L$		& 25,000	& 1		& 25,000		& 1		& 14,000		& 6		\\
Effective Temperature (K)		& $T_{\rm eff}$	& 48,000	& 2		& 48,000		& 2		& 46,500		& 6		\\
Magnitudes				& $U$		& 8.67		& 3		& 8.57			& 5		& 6.76			& 7		\\
					& $B$		& 9.93		& 3		& 9.73			& 5		& 8.02			& 7		\\
					& $V$		& 10.17		& 3		& 10.01			& 5		& 8.29			& 7		\\
Distance (kpc)				& $d$		& $\simeq$ 2.7	& 2		& 2.0$^{+0.9}_{-0.6}$	& 4		& 0.65 $\pm$ 0.10	& 8		\\
Terminal wind velocity (km s$^{-1}$)	& $v_{\infty}$	& 2,000		& 2		& 2,000			& 2		& 1,350			& 9		\\
Mass--loss rate ($M_{\odot}$ yr$^{-1}$)	& $\dot M$	& 10$^{-8.2}$	& 2		& 10$^{-8.5}$		& 2		& 10$^{-8.5}$		& 6		\\ \hline
\end{tabular}
\end{center}
\begin{small}
References: 1 - \citet{Darius+79}; 2 - \citet{JefferyHamann10}; 3 - \citet{Jordi+91}; 4 - \citet{BauerHusfeld95}; 5 - \citet{Landolt73}; 6 - \citet{Hamann10}; 7 - \citet{LandoltUomoto07}; 8 - \citet{KudritzkiSimon78}; 9 - \citet{Hamann+81}
\end{small}
\end{table*}

Among the hot subdwarf (sd) stars, which are evolved He-core burning low-mass stars \citep{Heber09}, the sdO stars are those which show the highest temperatures (T$_{\rm eff}>$ 40 kK). Apart from this characteristic, sdO stars (sdOs) are characterized by a wide range of values for the surface gravity (log($g$) = 4--6.5) and helium abundance (-3.5 $\lsim$ log($n_{\rm He}n_{\rm H}^{-1}$) $\lsim$ 3). Indeed they form a rather heterogeneus class of stars, which includes both He-poor and He-rich stars \citep{HeberJeffery92,Heber+06,Hirsch+08}, and `luminous' and `compact' stars, according to their low or high surface gravity, respectively \citep{Napiwotzki08}. This variety of properties is probably the consequence of different evolutionary histories \citep{Heber09,Geier15}: in the case of the compact stars, the He-poor ones are post-EHB stars, while the origin of the He-rich ones might be either the merging of two He-core or C/O-core white dwarfs \citep{Iben90,SaioJeffery00,SaioJeffery02} or the so-called late hot-flasher scenario \citep{Brown+01}; instead the luminous sdO stars are post-AGB stars.
Evolutionary models suggest that most sdO stars are the outcome of the evolution of single stars, but some of them could descend from binary systems which underwent a common-envelope phase; in this case it is possible that the sdO stars has a compact companion, typically a white dwarf (WD).

Up to now, sdO stars have been deeply investigated in the optical/UV domain, where several of them are rather bright; on the other hand, only few of them are known as X-ray sources. In the case of the binary \HD, the detection of pulsed ($P$ = 13.18 s) soft X-rays \citep{Israel+97} indicates that this emission originates from accretion onto a compact object, most likely a massive white dwarf \citep{Mereghetti+09}. For this binary we detected an evident X-ray emission also when the compact companion is eclipsed by the sdO star, suggesting the possibility of intrinsic X-ray emission of the sdO star \citep{Mereghetti+13}. Another sdO star recently detected at X-rays is \BDone: the \XMM\ observation of this He-rich star revealed soft X-ray emission, with a spectrum similar to that of \HD, and a possible periodicity of 19.16 s (at 3 $\sigma$ confidence level), which suggests that also \BDone\ has a compact companion \citep{LaPalombara+12}. In order to enlarge the sample of sdO stars observed at X-rays, we performed with \Chandra\ HRC-I a survey of a complete flux-limited sample of sdO stars and discovered three additional X-ray emitting stars \citep{LaPalombara+14}: the luminous and He-rich sdO star \BDtwo\ \citep{JefferyHamann10} and the compact (log($g$) $>$ 6) and He-poor stars Feige 34 and \BDthree\ \citep{Thejll+91,ZaninWeinberger97}.

In this paper we report on the results of a follow-up observation of \BDtwo, performed with \XMM, which allowed us to investigate in detail the spectral and timing properties of the X-ray emission discovered with \Chandra. This star was identified as an sdO star by \citet{Wolff+74}, who detected several emission He lines but no H lines in its blue spectrum. Their spectroscopic analysis gave a surface gravity log$g \lsim$ 4.5 and a temperature $T \lsim$ 50 kK; comparable values ($T \simeq$ 55 kK and log$g \simeq$ 4.0) were estimated from the low resolution IUE spectrum \citep{Darius+79}, which also gave an estimate of the star luminosity (log($L_{\rm bol}/L_{\odot}$ = 4.4). There is no evidence for a compact companion for \BDtwo. The possible detection of an infrared excess (at 2-$\sigma$ confidence level, \citet{Ulla+98}), if confirmed, could imply a companion star of spectral type earlier than G4. Based on high-resolution ultraviolet and optical spectra and on UV-optical-IR photometry, and using the latest generation of models for spherically expanding stellar atmospheres, \citet{JefferyHamann10} found that \BDtwo\ is characterized by a significant mass-loss rate ($\dot M = 10^{-8.2} M_{\odot}$ y$^{-1}$). Their analysis yielded revised and better constrained spectral parameters, very similar to those of \BDone, the other X-ray detected He-rich sdO star. The best-fit distance modulus obtained from this analysis is \textit{DM} = 12.2, corresponding to a distance of $\simeq$ 2.7 kpc. Our observation of \BDtwo\ with \Chandra\ HRC-I provided a detection with a count rate CR = 3.6$^{+1.1}_{-0.9}$ cts s$^{-1}$, which, assuming a spectrum similar to that of \HD\ and \BDone, implies an X-ray luminosity $L_{\rm X} \sim 10^{31}$ erg s$^{-1}$. For comparison, in Table~\ref{parameters} we report the main parameters of the three luminous sdO stars detected in X-rays.


\begin{table*}[!t]
\caption{List of observations of \BDtwo\ performed by \XMM.}\label{observations}
\begin{center}
\begin{tabular}{ccccc} \hline \hline
Revolution	& Observation ID	& Start time			& \multicolumn{2}{c}{Effective exposure}	\\
		& 			& (YYYY-MM-DD\@~hh:mm:ss)	& \pn\ (ks)	& MOS (ks)			\\ \hline
2627		& 0740140301		& 2014-04-14@15:00:25		& 4.2		& 8.5				\\
2629		& 0740140501		& 2014-04-18@14:43:21		& 7.0		& 11.3				\\
2630		& 0740140401		& 2014-04-20@15:45:37		& 7.9		& 10.0				\\
2631		& 0740140601		& 2014-04-22@14:36:30		& 9.9		& 14.4				\\
2632		& 0740140701		& 2014-04-24@14:06:44		& 5.5		& 6.8				\\ \hline
\end{tabular}
\end{center}
\end{table*}

\section{Observations and data analysis}\label{data}

\BDtwo~was observed with \XMM\ in April 2014. At that time the source was visible only for the first $\sim$ 20 ks of each \XMM\ orbit: therefore, five different observations were performed, between April 14 and 24 (see Table~\ref{observations}). The three EPIC cameras, i.e.~one \pn\ \citep{Struder+01} and two MOS \citep{Turner+01}, were always operated in \textit{full frame} mode, with time resolution of 73 ms for the \pn\ and of 2.6 s for the two MOS cameras; taking into account all the observations, the total effective exposure time was, respectively, of $\simeq$ 34.5 ks and $\simeq$ 50 ks. For all cameras the medium thickness filter was used.

\begin{figure*}[]
\centering
\includegraphics[angle=0,width=18truecm]{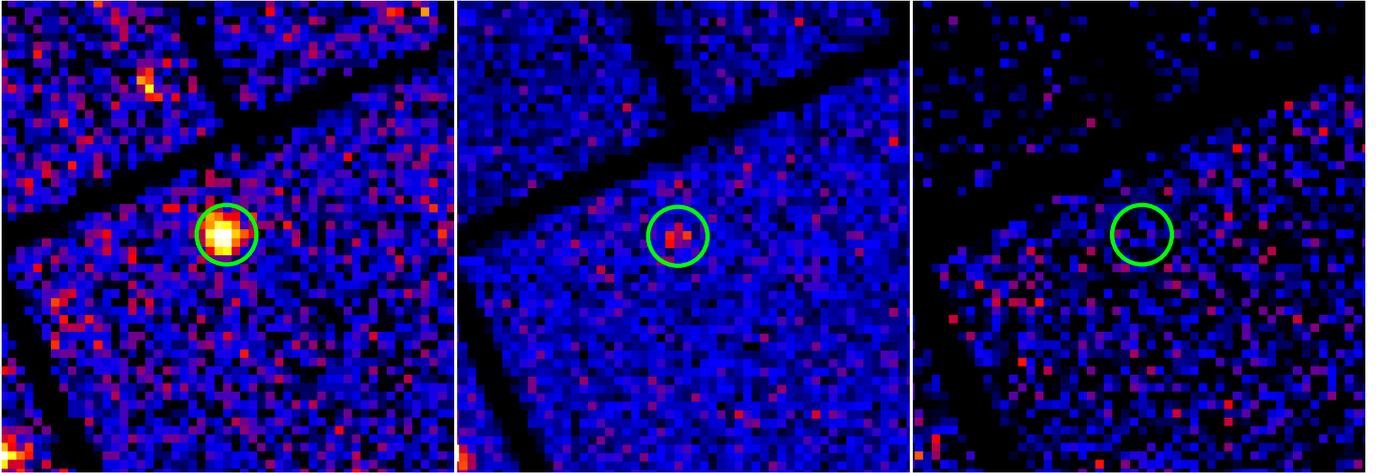}
\caption{EPIC \pn\ image of the sky region around \BDtwo\ in the energies ranges 0.15-0.5 (\textit{left}), 0.5-1.5 (\textit{centre}), and 1.5-10 keV (\textit{right}). The green circle (15$''$ radius) indicates the source position.}
\label{image}
\end{figure*}

We used version 13.5 of the \XMM~{\em Science Analysis System} (\texttt{SAS}) to process the event files. For the data analysis we selected only the events with pattern in the range 0--4 (i.e.~mono-- and bi--pixel events) for the \pn\ camera and 0--12 (i.e. from 1 to 4 pixel events) for the two MOS. For each camera, we merged together the data of the five observations and accumulated the images in various energy ranges. We found that \BDtwo\ is significantly detected at the coordinates R.A. = 9$^h$ 24$^m$ 26.4$^s$, Dec. = +36$^\circ$ 42$'$ 52.8$''$, which differ by 0.7$''$ from the position of \BDtwo. This difference is consistent with the $\sim$ 2$''$ r.m.s. astrometric accuracy of \XMM\footnote{http://xmm2.esac.esa.int/docs/documents/CAL-TN-0018.ps.gz}. In each of the five observations a point source is clearly detected below 0.5 keV, while considering the five observation merged together the source is detected up to $\sim$ 1.5 keV (Fig.~\ref{image}). This implies that the source spectrum is very soft. All the observations were partly affected by high instrumental background. However, since the spectrum of the instrumental background is rather hard, the background contamination has a limited impact on the source spectral analysis: therefore, we considered the whole data set, without rejecting the time intervals with the highest particle background. The source net count rate in the 0.15--1.5 keV range is (1.7 $\pm$ 0.2)$\times$10$^{-2}$ cts s$^{-1}$ and (2.5 $\pm$ 0.3)$\times$10$^{-3}$ cts s$^{-1}$ for the \pn\ and each of the two MOS, respectively.

For the timing and spectral analysis, we used the data of the whole observation and the three EPIC cameras; we extracted the source events from a circular region with radius 15$''$ centered at the source position, while the corresponding background events were accumulated on circular areas free of sources and radii of 30$''$ and 120$''$ for the \pn\ and the two MOS cameras, respectively. We converted the arrival times to the solar system barycenter, then we combined the three datasets in a single event list. The background--subtracted light curve of \BDtwo\ does not show any variability on time scales from hundreds of seconds to the observation length. We looked for possible periodicities in the X-ray emission, but we found no evidence of periodic signals; this search was unsuccessful not only for the five individual observations, but also when considering together all of them. In all cases we estimated an upper limit of $\sim$ 30 \% on the pulsed fraction, for a sinusoidal modulation between 1 and 5000 s.

For the spectral analysis we considered first the \pn\ data, since the source soft spectrum and the lower sensitivity of the MOS cameras at low energies strongly reduced the count statistics. We verified that the addition of the MOS data gave consistent results. We generated the response matrix and ancillary file using the \texttt{SAS} tasks \texttt{rmfgen} and \texttt{arfgen}. To ensure the applicability of the $\chi^{2}$ statistics, the spectrum was rebinned with a minimum of 30 net counts per bin; then, we fitted them using \texttt{XSPEC} (V 12.7.0). We only used the energy range 0.2--1.5 keV since at higher energies the background dominates and the source flux is negligible. In the following, all the spectral uncertainties and upper limits are given at the 90 \% confidence level for one interesting parameter, and we assume a source distance of 2.7 kpc \citep{JefferyHamann10}; we adopted the results of \citet{AndersGrevesse89} for the solar abundances of the atomic elements.

The source spectrum is very soft and we tried to describe it with different models (see Table~\ref{spectral_fit}). The fit with an absorbed power law (PL) is formally acceptable ($\chi^{2}_{\nu} <$ 2) but gives a very large and unrealistic photon index ($\Gamma \simeq$ 5), while a fit with a blackbody model is rejected by the data ($\chi^{2}_{\nu} >$ 2). 
An absorbed power law plus blackbody gives a good fit, but with unrealistic (for the power-law photon index) or unconstrained (for the blackbody normalization) values of the model parameters.
We note that while this model is physically motivated for \BDone, where the observed X-ray flux can be attributed to accretion onto a compact companion, this is not the case for \BDtwo, for which no evidence of a compact companion has been found. For this reason,
we consider in the following the possibility that the X-ray emission detected in \BDtwo\ has the same origin of that observed in the normal, giant and supergiant early-type O stars.

\begin{figure}[h]
\centering
\resizebox{\hsize}{!}{\includegraphics[angle=-90,clip=true]{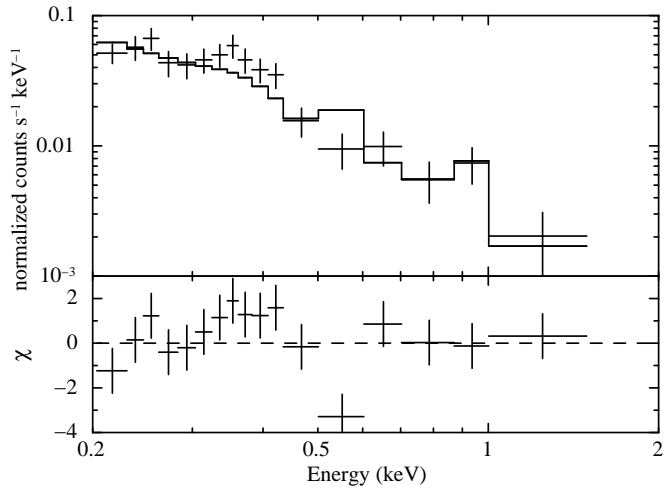}}
\caption{\textit{Top panel}: EPIC \pn\ spectrum of \BDtwo\ with the best--fit model composed by the sum of two thermal plasma components, with solar abundances. \textit{Bottom panel}: residuals (in units of $\sigma$) between data and model.}
\label{spectrum1}
\end{figure}

For a large sample of this type of stars observed with \XMM\ the spectrum can be described by the sum of different thermal plasma components (\texttt{MEKAL} in \texttt{XSPEC}), with temperatures between $\simeq$ 0.1 and $\simeq$ 5 keV \citep{Naze09}. Therefore, we tried to use the same approach also in the case of \BDtwo. We clearly found that, assuming solar abundances, with this model it was not possible to obtain an acceptable fit, even if we considered the sum of two \texttt{MEKAL} components at different temperatures (Fig.~\ref{spectrum1} and Table~\ref{spectral_fit}). Therefore, we modified the model abundances by taking into account the values of the single chemical elements considered by \citet{JefferyHamann10}. Since there are no abundance measurements for \BDtwo, they adopted the same overabundance of He, C, N, Si, and Fe obtained by \citet{BauerHusfeld95} for \BDone, which is very similar from the spectroscopic point of view.

\begin{figure}[h]
\centering
\resizebox{\hsize}{!}{\includegraphics[angle=-90,clip=true]{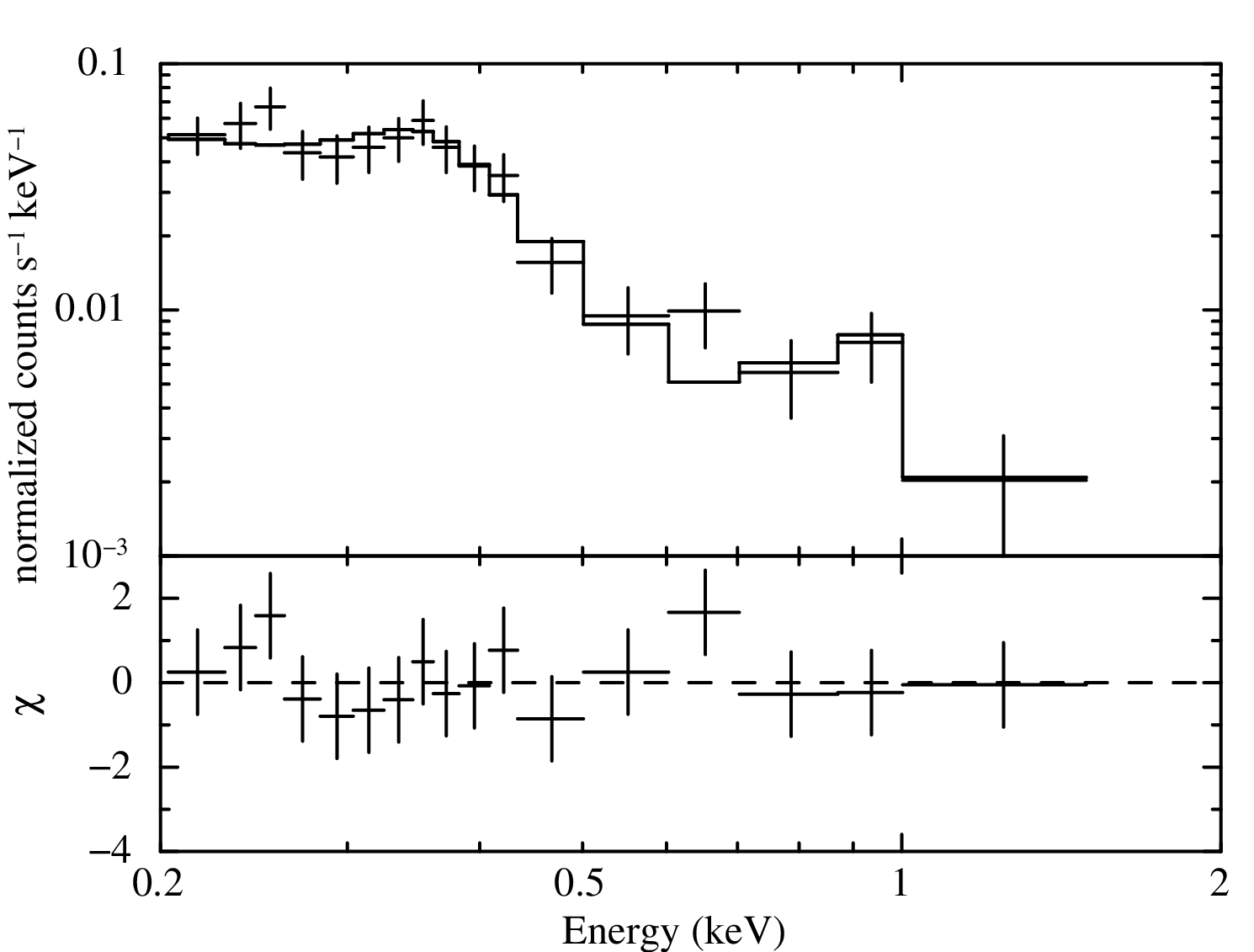}}
\caption{\textit{Top panel}: EPIC \pn\ spectrum of \BDtwo\ with the best--fit model composed by the sum of two thermal plasma components, with abundances from \citet{JefferyHamann10}. \textit{Bottom panel}: residuals (in units of $\sigma$) between data and model.}
\label{spectrum2}
\end{figure}

\begin{table}[!t]
\caption{Summary of the best-fit parameters of \BDtwo\ obtained with different spectral models.}\label{spectral_fit}
\begin{center}
\begin{tabular}{ccc} \hline \hline
Parameter	& Unit		& Value			\\ \hline
\multicolumn{3}{c}{Power law}	\\
$N_{\rm H}$	& cm$^{-2}$	& (3.5$^{+4.0}_{-2.7}$)$\times10^{20}$ \\
$\Gamma$	& -	& 5.0$^{+1.7}_{-1.1}$ \\
$\chi^{2}_{\nu}$	& -	& 1.76	\\
Degrees of freedom	& -	& 14	\\ \hline
\multicolumn{3}{c}{Blackbody}	\\
$N_{\rm H}$	& cm$^{-2}$	& (0.9$^{+5.8}_{-0.9}$)$\times10^{20}$ \\
$kT$		& eV		& 67$^{+12}_{-22}$ \\
$\chi^{2}_{\nu}$	& -	& 2.46	\\
Degrees of freedom	& -	& 14	\\ \hline
\multicolumn{3}{c}{Power law + Blackbody}	\\
$N_{\rm H}$	& cm$^{-2}$	& (2.0$^{+2.0}_{-1.3}$)$\times10^{21}$ \\
$\Gamma_{\rm PL}$	& -	& 3.9$^{+2.3}_{-1.7}$ \\
$kT_{\rm BB}$ 	& eV	& 22$^{+16}_{-12}$ \\
$\chi^{2}_{\nu}$	& -	& 0.76	\\
Degrees of freedom	& -	& 12	\\ \hline
\multicolumn{3}{c}{Mekal + Mekal}	\\
\multicolumn{3}{c}{(with solar abundances)}	\\
$N_{\rm H}$	& cm$^{-2}$	& (1.8$\pm$1.2)$\times10^{20}$	\\
$kT_{1}$ 	& eV		& 81$^{+5}_{-0}$ \\
$kT_{2}$ 	& eV	& 800$^{+330}_{-230}$ \\
$\chi^{2}_{\nu}$	& -	& 2.15	\\
Degrees of freedom	& -	& 12	\\ \hline
\multicolumn{3}{c}{Mekal + Mekal}	\\
\multicolumn{3}{c}{(with abundances from \citet{JefferyHamann10})}	\\
$N_{\rm H}$	& cm$^{-2}$	& 1$\times10^{20}$ (fixed) \\
$kT_{1}$ 	& eV		& 120$\pm$30 \\
$kT_{2}$ 	& eV		& 840$^{+350}_{-210}$ \\
$\chi^{2}_{\nu}$	& -	& 0.71	\\
Degrees of freedom	& -	& 13	\\ \hline
\multicolumn{3}{c}{Mekal + Mekal}	\\
\multicolumn{3}{c}{(with abundances from \citet{JefferyHamann10})}	\\
\multicolumn{3}{c}{(with free Oxygen abundance)}	\\
$N_{\rm H}$	& cm$^{-2}$	& 1$\times10^{20}$ (fixed) \\
$kT_{1}$ 	& eV		& 100$^{+40}_{-20}$ \\
$kT_{2}$ 	& eV		& 840$^{+570}_{-250}$ \\
Oxygen Abundance	& -	& 310$\pm$250 \\
$\chi^{2}_{\nu}$	& -	& 0.64	\\
Degrees of freedom	& -	& 12	\\ \hline
\end{tabular}
\end{center}
\end{table}

\begin{figure}[h]
\centering
\resizebox{\hsize}{!}{\includegraphics[angle=-90,clip=true]{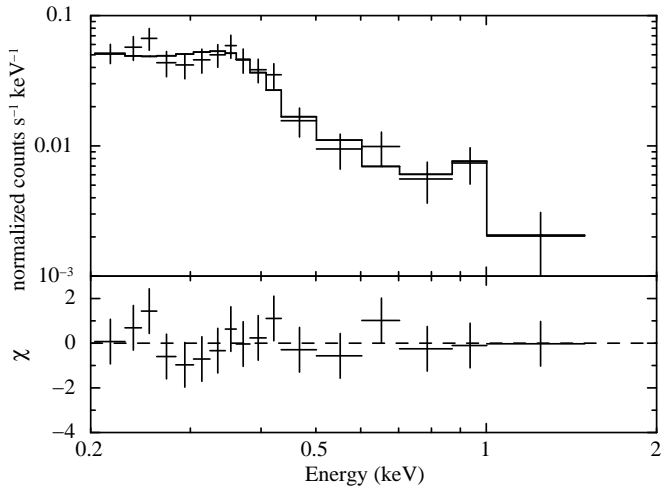}}
\caption{\textit{Top panel}: EPIC \pn\ spectrum of \BDtwo\ with the best--fit model composed by the sum of two thermal plasma components, with abundances from \citet{JefferyHamann10} and free Oxygen abundance. \textit{Bottom panel}: residuals (in units of $\sigma$) between data and model.}
\label{spectrum3}
\end{figure}

Assuming these abundances, a single thermal component provides an acceptable fit ($\chi^{2}_{\nu} <$ 2) but leaves large residuals at the high energies. Hence we considered a model composed by the sum of two absorbed components, at different temperatures. We checked that the estimated interstellar absorption is negligible and consistent with 0. Therefore we fixed it at $N_{\rm H} = 10^{20}$ cm$^{-2}$, which is the total absorption value across the Galaxy in the direction of \BDtwo. In this way we found a good fit ($\chi^{2}_{\nu}$ = 0.71 for 13 degrees of freedom, Fig.~\ref{spectrum2}) with $kT_{\rm 1}$ = 120$\pm$30 eV and $kT_{\rm 2}$ = 840$^{+350}_{-210}$ eV. The total flux in the energy range 0.2--1.5 keV is $f_{\rm X}$ = ($4.0^{+0.2}_{-0.3}$) $\times 10^{-14}$ erg cm$^{-2}$ s$^{-1}$; although 78 \% of the flux is due to the low-temperature component, the high-temperature component is significant at 3 $\sigma$ confidence level. The measured flux corresponds to a source luminosity $L_{\rm X}$ = ($3.3^{+0.2}_{-0.3}$) $\times 10^{31}$ erg s$^{-1}$. The fit leaves some residuals at $\sim$ 650 eV, thus suggesting the presence of the O\,{\sc viii} emission line (Fig.~\ref{spectrum2}).
This could be due to an underestimate of the real Oxygen abundance, since in our model we fixed it at the Solar value. Therefore we repeated the spectral fit with the same model but leaving the Oxygen abundance free to vary. In this way we obtained an improvement of the spectral fit (Fig.~\ref{spectrum3}) and we found that the best-fit abundance value is 310$\pm$250 times the Solar value: although it is not well constrained, this is consistent with the expected Oxygen overabundance in \BDtwo.


\section{Discussion}

The \XMM\ observation of \BDtwo\ enabled us to constrain the flux and spectrum of the X-ray emission recently discovered by \Chandra\ \citep{LaPalombara+14}. The measured flux $f_{\rm X} \simeq 4 \times 10^{-14}$ erg cm$^{-2}$ s$^{-1}$ confirms the estimate provided by the \Chandra\ detection. It implies a luminosity $L_{\rm X} \simeq 3.3 \times 10^{31}$ erg s$^{-1}$. Since the bolometric luminosity of \BDtwo\ is $L_{\rm bol} \simeq$ 25,000 $L_{\odot}$ \citep{Darius+79,JefferyHamann10}, the corresponding ratio is $L_{\rm X}/L_{\rm bol} \sim 10^{-6.5}$. This value is consistent with the `canonical' relation $L_{\rm X} \sim 10^{-7} \times L_{\rm bol}$ obtained for the normal, giant and supergiant early-type O stars, which are known since a long ago as X-ray sources \citep{Pallavicini+81,Sciortino+90,GuedelNaze09}. 
The  hypothesis of intrinsic origin for the X-ray emission of \BDtwo\ is further supported by the spectral analysis. In fact, considering the likely possibility of non-Solar composition, the  spectrum can be successfully described by the sum of two thermal plasma components,
as in normal early-type stars.




It is interesting to compare the properties of the three sdOs for which X-ray spectral information is available (Table~\ref{parameters2}).
The spectrum of \HD\ during the eclipse phase can be described by the sum of three thermal plasma components \citep{Mereghetti+13}. While the hottest component is required to account for the significant emission above $\sim$ 2 keV, the temperatures of the two coldest components are very similar to those of \BDtwo. The 0.2--10 keV   luminosity of \HD\ during the eclipse phase is $L_{\rm X} \simeq 3 \times 10^{30}$ erg s$^{-1}$, i.e. one order of magnitude lower than that of \BDtwo. However, due to its lower bolometric luminosity, also for \HD\ the X-ray/bolometric luminosity ratio is $L_{\rm X}/L_{\rm bol} \sim 10^{-7}$.
The spectrum of the other X-ray emitting sdO star, \BDone, can also be fit with a similar themal model with the He and metal abundances derived for this star \citep{JefferyHamann10} and the temperatures indicated in Table~\ref{parameters2}.
Its flux $f_{\rm X} \simeq 6.4 \times 10^{-14}$ erg cm$^{-2}$ s$^{-1}$ implies a luminosity $L_{\rm X} \simeq 2.9 \times 10^{31}$ erg s$^{-1}$ (for a  distance of 2 kpc), hence a X-ray/bolometric luminosity ratio $L_{\rm X}/L_{\rm bol}$ = $10^{-6.3}$.

These results indicate that the X-ray emission from these three luminous sdO stars is similar to that of normal O-type stars, which have luminosities up to a few $10^{33}$ erg s$^{-1}$. In these stars the X-ray emission
is due to turbulence and shocks in their strong embedded winds \citep{LucyWhite80,Owocki+88}.
Since the mass-loss rate in the radiation-driven winds of early type stars scales with the bolometric luminosity, and the X-ray emission originates in the stellar wind, a correlation between $L_{\rm X}$ and $L_{\rm bol}$ is not surprising (see, e.g., \citet{Owocki+13}).
In this respect it is interesting to note that the three X-ray emitting sdOs are among the few hot subdwarfs for which evidence of mass loss has been reported \citep{JefferyHamann10}. Our results indicate that, even if the winds of sdO stars are rather weak (e.g. $\dot M = 10^{-8.2} M_{\odot}$ yr$^{-1}$ for \BDtwo\ according to the estimate of \citet{JefferyHamann10}), they can produce X-ray emitting shocks, as in more luminous O type stars. In this framework, we note that our findings for sdO stars are supported by the methodology used and the results obtained by \citet{Cohen+14}, who investigated the X-ray spectra of O-type stars with very low mass-loss rates.

\begin{table*}[t]
\caption{Summary of the best-fit parameters of the three sdO stars observed with \XMM, when their spectrum is described with multi-temperature thermal-plasma components (\texttt{MEKAL} in \texttt{XSPEC}).}\label{parameters2}
\begin{center}
\begin{tabular}{cccccc} \hline \hline
Source			& $kT_{\rm 1}$			& $kT_{\rm 2}$			& $kT_{\rm 3}$	& log($L_{\rm X}/L_{\rm bol}$)	& Reference		\\
			& (keV)				& (keV)				& (keV)		&				&			\\ \hline
\HD\ (in eclipse)	& 0.13$\pm$0.02			& 0.71$_{-0.19}^{+0.15}$	& 5 (fix)	& -7.1				& \citet{Mereghetti+13}	\\
\BDone\			& 0.17$_{-0.01}^{+0.02}$	& 0.72$_{-0.10}^{+0.22}$	& -		& -6.3				& This work		\\
\BDtwo\			& 0.13$\pm$0.01			& 0.79$\pm$0.15			& -		& -6.5				& This work		\\ \hline
\end{tabular}
\end{center}
\end{table*}

\begin{figure}[h]
\centering
\resizebox{\hsize}{!}{\includegraphics[angle=-90,clip=true]{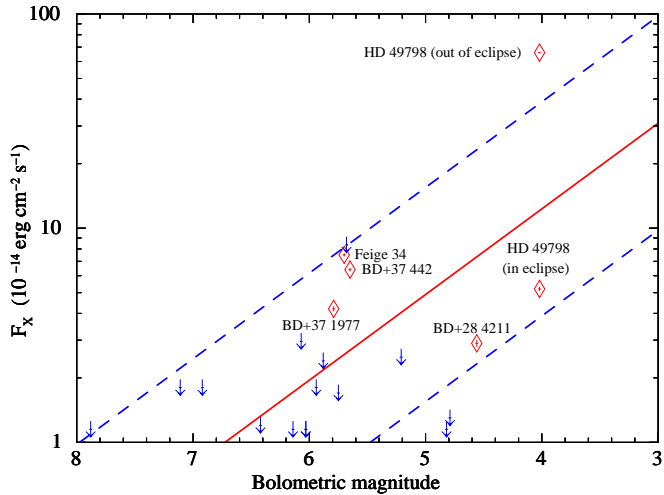}}
\caption{Level of the X-ray flux (or its upper limit for the undetected sources) of the sdO stars observed at X-rays, as a function of their bolometric magnitude. The upper and lower blue lines (corresponding to $f_{\rm X}/f_{\rm bol} = 10^{-6.2}$ and $f_{\rm X}/f_{\rm bol} = 10^{-7.2}$, respectively) include the range of expected values for the main-sequence early-type stars; the red line corresponds to $f_{\rm X}/f_{\rm bol} = 10^{-6.7}$, which is the best-fit relation found by \citet{Naze09} for this type of stars.}
\label{X-bol}
\end{figure}

The latest \Chandra\ detections of sdO stars reinforce the hypothesis that also this type of stars is a class of X-ray sources; in this respect, the detection not only of the luminous stars but also of the compact ones is very promising. In addition, we also note that in all cases the estimated $L_{\rm X}/L_{\rm bol}$ ratio agrees with that found in the heavier early-type stars: therefore also in the compact sdO stars the X-ray emission could be attributed to turbulence and shocks in their winds. In order to provide an overview of all the sdO stars observed at X-rays up to now, in Fig.~\ref{X-bol} we report, as a function of their bolometric magnitude, the X-ray flux of the detected stars and its upper limit for the undetected ones; for \HD, \BDone, and \BDtwo\ the flux value is based on the spectral fit provided by the \XMM\ data, while for the other sources it is based on the count rate value or limit provided by \Chandra\ HRC-I, assuming an emission spectrum similar to that of \BDtwo. For comparison, the two dashed lines trace the region corresponding to the typical X-ray-to-bolometric flux ratio for the normal O-type stars. The plot shows that almost all the observed stars are within this region, therefore the possible presence of intrinsic X-ray emission also from the stars undetected so far cannot be excluded.

\acknowledgements
This work is based on observations obtained with \XMM, an ESA science mission with instruments and contributions directly funded by ESA Member States and NASA. PE acknowledges a Fulbright Research Scholar grant administered by the U.S.--Italy Fulbright Commission and is grateful to the Harvard--Smithsonian Center for Astrophysics for hosting him during his Fulbright exchange.

\bibliographystyle{aa}
\bibliography{biblio}

\begin{thebibliography}{41}
\expandafter\ifx\csname natexlab\endcsname\relax\def\natexlab#1{#1}\fi

\bibitem[Anders \& Grevesse(1989)]{AndersGrevesse89} Anders, E., \& Grevesse, N.\ 1989, \gca, 53, 197 

\bibitem[{{Bauer} \& {Husfeld}(1995)}]{BauerHusfeld95} {Bauer}, F. \& {Husfeld}, D. 1995, \aap, 300, 481

\bibitem[Brown et al.(2001)]{Brown+01} Brown, T.~M., Sweigart, A.~V., Lanz, T., Landsman, W.~B., \& Hubeny, I.\ 2001, \apj, 562, 368

\bibitem[Cohen et al.(2014)]{Cohen+14} Cohen, D.~H., Wollman, E.~E., Leutenegger, M.~A., et al.\ 2014, \mnras, 439, 908

\bibitem[Darius et al.(1979)]{Darius+79} Darius, J., Giddings, J.~R., \& Wilson, R.\ 1979, The First Year of IUE, 363

\bibitem[{Geier}(2015)]{Geier15} {Geier}, S. 2015, arXiv:1503.01625 

\bibitem[G{\"u}del \& Naz{\'e}(2009)]{GuedelNaze09} G{\"u}del, M., \& Naz{\'e}, Y.\ 2009, \aapr, 17, 309 

\bibitem[{{Hamann} {et~al.}(1981){Hamann}, {Gruschinske}, {Kudritzki}, \& {Simon}}]{Hamann+81} {Hamann}, W., {Gruschinske}, J., {Kudritzki}, R.~P., \& {Simon}, K.~P. 1981, \aap, 104, 249

\bibitem[{{Hamann}(2010)}]{Hamann10} {Hamann}, W. 2010, \apss, 119

\bibitem[{{Heber} \& {Jeffery}(1992)}]{HeberJeffery92} {Heber}, U. \& {Jeffery}, C.~S., eds. 1992, Lecture Notes in Physics, Berlin Springer Verlag, Vol. 401, {The Atmospheres of Early-Type Stars}

\bibitem[{{Heber} {et~al.}(2006){Heber}, {Hirsch}, {Str{\"o}er}, \& {et al.}}]{Heber+06} {Heber}, U., {Hirsch}, H., {Str{\"o}er}, A., \& {et al.} 2006, Baltic Astronomy, 15, 91

\bibitem[{{Heber}(2009)}]{Heber09} {Heber}, U. 2009, \araa, 47, 211

\bibitem[Heber et al.(2014)]{Heber+14} Heber, U., Geier, S., Irrgang, A., et al.\ 2014, 6th Meeting on Hot Subdwarf Stars and Related Objects, 481, 307 

\bibitem[{{Hirsch} {et~al.}(2008){Hirsch}, {Heber}, \& {O'Toole}}]{Hirsch+08} {Hirsch}, H.~A., {Heber}, U., \& {O'Toole}, S.~J. 2008, in Astronomical Society of the Pacific Conference Series, Vol. 392, Hot Subdwarf Stars and Related Objects, ed. {U.~Heber, C.~S.~Jeffery, \& R.~Napiwotzki}, 131

\bibitem[Iben(1990)]{Iben90} Iben, I., Jr.\ 1990, \apj, 353, 215

\bibitem[Israel et al.(1997)]{Israel+97} Israel, G.~L., Stella, L., Angelini, L., et al.\ 1997, \apjl, 474, L53 

\bibitem[{{Jeffery} \& {Hamann}(2010)}]{JefferyHamann10} {Jeffery}, C.~S. \& {Hamann}, W.-R. 2010, \mnras, 404, 1698

\bibitem[Jordi et al.(1991)]{Jordi+91} Jordi, C., Figueras, F., Paredes, J.~M., Rossello, G., \& Torra, J.\ 1991, \aaps, 87, 229 

\bibitem[{{Kudritzki} \& {Simon}(1978)}]{KudritzkiSimon78} {Kudritzki}, R.~P. \& {Simon}, K.~P. 1978, \aap, 70, 653

\bibitem[{{Landolt}(1973)}]{Landolt73} {Landolt}, A.~U. 1973, \pasp, 85, 661

\bibitem[{{Landolt} \& {Uomoto}(2007)}]{LandoltUomoto07} {Landolt}, A.~U. \& {Uomoto}, A.~K. 2007, \aj, 133, 768

\bibitem[La Palombara et al.(2012)]{LaPalombara+12} La Palombara, N., Mereghetti, S., Tiengo, A., \& Esposito, P.\ 2012, \apjl, 750, L34 

\bibitem[La Palombara et al.(2014)]{LaPalombara+14} La Palombara, N., Esposito, P., Mereghetti, S., \& Tiengo, A.\ 2014, \aap, 566, AA4 

\bibitem[Lucy \& White(1980)]{LucyWhite80} Lucy, L.~B., \& White, R.~L.\ 1980, \apj, 241, 300 

\bibitem[{{Mereghetti} {et~al.}(2009){Mereghetti}, {Tiengo}, {Esposito}, \& {et  al.}}]{Mereghetti+09} {Mereghetti}, S., {Tiengo}, A., {Esposito}, P., \& {et al.} 2009, Science, 325,  1222

\bibitem[Mereghetti et al.(2013)]{Mereghetti+13} Mereghetti, S., La Palombara, N., Tiengo, A., et al.\ 2013, \aap, 553, A46 

\bibitem[{{Napiwotzki}(2008)}]{Napiwotzki08} {Napiwotzki}, R. 2008, in Astronomical Society of the Pacific Conference Series, Vol. 392, Hot Subdwarf Stars and Related Objects, ed. {U.~Heber, C.~S.~Jeffery, \& R.~Napiwotzki}, 139

\bibitem[Naz{\'e}(2009)]{Naze09} Naz{\'e}, Y.\ 2009, \aap, 506, 1055

\bibitem[Owocki et al.(1988)]{Owocki+88} Owocki, S.~P., Castor, J.~I., \& Rybicki, G.~B.\ 1988, \apj, 335, 914

\bibitem[Owocki et al.(2013)]{Owocki+13} Owocki, S.~P., Sundqvist, J.~O., Cohen, D.~H., \& Gayley, K.~G.\ 2013, \mnras, 429, 3379

\bibitem[Pallavicini et al.(1981)]{Pallavicini+81} Pallavicini, R., Golub, L., Rosner, R., et al.\ 1981, \apj, 248, 279 

\bibitem[Saio \& Jeffery(2000)]{SaioJeffery00} Saio, H., \& Jeffery, C.~S.\ 2000, \mnras, 313, 671 

\bibitem[Saio \& Jeffery(2002)]{SaioJeffery02} Saio, H., \& Jeffery, C.~S.\ 2002, \mnras, 333, 121

\bibitem[Sciortino et al.(1990)]{Sciortino+90} Sciortino, S., Vaiana, G.~S., Harnden, F.~R., Jr., et al.\ 1990, \apj, 361, 621 

\bibitem[{{Str{\"u}der} {et~al.}(2001){Str{\"u}der}, {Briel}, {Dennerl}, \& {et al.}}]{Struder+01} {Str{\"u}der}, L., {Briel}, U., {Dennerl}, K., \& {et al.} 2001, \aap, 365, L18

\bibitem[Thejll et al.(1991)]{Thejll+91} Thejll, P., MacDonald, J., \& Saffer, R.\ 1991, \aap, 248, 448 

\bibitem[{{Turner} {et~al.}(2001){Turner}, {Abbey}, {Arnaud}, \& {et al.}}]{Turner+01} {Turner}, M.~J.~L., {Abbey}, A., {Arnaud}, M., \& {et al.} 2001, \aap, 365, L27

\bibitem[Ulla \& Thejll(1998)]{Ulla+98} Ulla, A., \& Thejll, P.\ 1998, \aaps, 132, 1 

\bibitem[Wolff et al.(1974)]{Wolff+74} Wolff, S.~C., Pilachowski, C.~A., \& Wolstencroft, R.~D.\ 1974, \apjl, 194, L83 

\bibitem[Zanin \& Weinberger(1997)]{ZaninWeinberger97} Zanin, C., \& Weinberger, R.\ 1997, Planetary Nebulae, 180, 290 

\end{thebibliography}

\end{document}